\documentclass[reprint,amsmath,amssymb,aps]{revtex4-1}
\usepackage{graphicx}
\usepackage{color}
\usepackage{amsmath}
\usepackage{amssymb}
\usepackage{amsthm}
\usepackage{booktabs}
\usepackage{float}
\usepackage[dvipdfm,colorlinks,urlcolor=blue,linkcolor=blue,anchorcolor=blue,citecolor=blue]{hyperref}
\usepackage{lineno}

\begin{document}

\title{Nonintegrability-driven Transition from Kinetics to Hydrodynamics}

\author{Weicheng Fu$^{1,3}$}
\thanks{These authors contributed equally to this work.}
\author{Zhen Wang$^{2}$}
\thanks{These authors contributed equally to this work.}
\author{Yisen Wang$^{3}$}
\author{Yong Zhang$^{4,3}$}
\email{yzhang75@xmu.edu.cn}
\author{Hong Zhao$^{4,3}$}
\email{zhaoh@xmu.edu.cn}

\affiliation{
$^1$ Department of Physics, Tianshui Normal University, Tianshui 741001, Gansu, China\\
$^2$ CAS Key Laboratory of Theoretical Physics and Institute of Theoretical Physics,
Chinese Academy of Sciences, Beijing 100190, China\\
$^3$ Lanzhou Center for Theoretical Physics, Key Laboratory of Theoretical Physics of Gansu Province, Lanzhou University, Lanzhou, Gansu 730000, China\\
$^4$ Department of Physics, Xiamen University, Xiamen 361005, Fujian, China
}

\date{\today }

\begin{abstract}
Nonintegrability plays a crucial role in  thermalization and transport processes in many-body Hamiltonian systems, yet its quantitative effects remain unclear. To reveal the connection between the macroscopic relaxation properties and the underlying dynamics, the one-dimensional diatomic hard-point model as an illustrating example  was studied analytically and numerically. We demonstrate how the system transitions from kinetic behavior to hydrodynamic behavior as the nonintegrability strength increases. Specifically, for the thermalization dynamics, we find a power-law relationship between the thermalization time and the perturbation strength near integrable regime, whereas in the far from integrable regime, the hydrodynamics dominates and the thermalization time becomes independent of the perturbation strength and exhibits a strong size-dependent behavior. Regarding transport behavior, our results further establish a threshold for the nonintegrable strength of this transition. Consequently, we can predict which behavior dominates the transport properties of the system. Especially, an explicit expression of the thermal conductivity contributed by the kinetics is given. Finally, possible applications were briefly discussed.
\end{abstract}

\maketitle

\emph{Introduction.}--The study on behaviors of relaxation and transport of many-body Hamiltonian systems is a core subject in non-equilibrium statistical physics. The one-dimensional (1D) diatomic hard-point (DHP) model is widely used to investigate the related fundamental problems, such as the ergodicity hypothesis \cite{casati1976computer}, transport phenomena \cite{casati1986energy,PhysRevE.67.015203,PhysRevE.89.022111,Li2004,PhysRevLett.110.070604,PhysRevLett.121.080602},  local equilibrium state \cite{PhysRevLett.86.3554}, and the Boltzmann $H$-theorem \cite{PhysRevE.84.031127}. In recent years, extensive research has focused on validating the Fourier heat conduction law, and a consensus has gradually emerged that this model exhibits divergent heat conduction \cite{PhysRevLett.89.180601,PhysRevLett.94.244301,lepri2003thermal,Dhar08AdvPhys}. This anomalous transport behavior has been addressed within the framework of hydrodynamics \cite{PhysRevLett.89.200601,PhysRevLett.108.180601,PhysRevLett.111.230601}, which suggests that three conservation quantities of the system, i.e., energy, momentum, and particle number, govern the relaxation process of fluctuations, resulting in power-law decay. However, there is also numerical evidence of deviations from hydrodynamic behavior, where under certain parameters, the fluctuations of the system exhibit exponential decay, thereby showing normal transport behavior \cite{PhysRevE.90.032134}. This is a typical kinetic behavior, and the relaxation of system fluctuations is described by the standard Boltzmann equation. Based on this, a new understanding has emerged, suggesting a transition from kinetic behavior to hydrodynamic behavior in the system, which occurs at some time or size threshold \cite{PhysRevE.97.010103,PhysRevLett.125.040604}. However, whether and how this transition depends on the underlying dynamics, such as nonintegrability, remains unclear. Another fundamental question is how this transition affects the thermalization dynamics of the system.

To address these issues, we separately investigated the thermalization dynamics and transport behavior of the 1D DHP model \cite{casati1986energy}. Through analytical analysis, confirmed by numerical simulations, we revealed how the system transitions from kinetic behavior to hydrodynamic behavior as the strength of nonintegrability increases. For thermalization dynamics, we studied the process of energy equipartition. We found that in the vicinity of integrability, the thermalization time exhibits a power-law dependence on the perturbation strength, whereas far from integrability, the dynamics of hydrodynamics dominates, and the thermalization time becomes independent of the perturbation strength, displaying strong size effects. Regarding transport behavior, our results quantitatively determined how this transition depends on non-integrability and the transition threshold. Our findings can predict the parameter ranges that lead to kinetic-dominated or hydrodynamic-dominated behavior.

\emph{Setup.}--We consider a 1D DHP system which consists of $N$ particles with alternative masses $m_1=1-{\delta}/{2}$ and $m_2=1+{\delta}/{2}$, where $\delta\in[0,2)$ is mass difference, thus the mass density is unity. Besides, we set the number density of particles is unity as well, i.e., the system size $L=N$. Therefore, the sound speed in this system has a concise form \cite{wong2002handbook}
\begin{equation}\label{eq-cs}
  c_{\rm s} = \sqrt{\gamma k_{\rm B} T}=\sqrt{3T},
\end{equation}
where $\gamma$ is the heat capacity ratio (here $\gamma=3$), $k_{\rm B}$ is the Boltzmann's constant which is set to be unity throughout, and $T$ is the temperature.

The particles move freely except for elastic collisions with the nearest neighbors. Suppose the $i$th particle and the $(i+1)$th particle undergo an elastic collision, the velocities evolve as
\begin{equation}
\begin{bmatrix}
  \tilde{v}_i \\
  \tilde{v}_{i+1}
\end{bmatrix}=
\begin{bmatrix}
 \frac{m_i-m_{i+1}}{m_i+m_{i+1}} & \frac{2m_{i+1}}{m_i+m_{i+1}}\\
 \frac{2m_{i}}{m_i+m_{i+1}} & \frac{m_{i+1}-m_i}{m_i+m_{i+1}}
\end{bmatrix}
\begin{bmatrix}
  v_i \\
  v_{i+1}
\end{bmatrix},
\end{equation}
where $v_i$ and $\tilde{v}_i$ denote the velocities before and after the collision, respectively.

Consider that two neighbouring particles collide if and only if their relative velocity $\nu>0$, so the average time for once collision is
\begin{equation}\label{eq-meanTime}
\theta=\frac{a}{\int_{0}^{\infty}\nu f(\nu)d\nu}
  =\sqrt{\frac{\pi(1-\delta^2/4)}{T}},%\frac{1}{2}\sqrt{\frac{\pi(4-\delta^2)}{T}}
\end{equation}
where $a=L/N=1$ is the mean spacing of particles, and $f(\nu)=\frac{1}{\sigma\sqrt{2\pi}}e^{-\frac{\nu^2}{2\sigma^2}}$ is the distribution function, where $\sigma=\sqrt{{2T}/{(1-\delta^2/4)}}$. In fact, $\theta$ is the mean time of free motion, i.e., the characteristic time of kinetic stage \cite{bogoliubov1962problems}. From Eq. (\ref{eq-meanTime}), we have $\theta\to \sqrt{{\pi}/{T}}$ for small $\delta$; and $\theta\to \sqrt{{\pi(2-\delta)}/{T}}$ for $\delta\to2$.

\emph{Thermalization.}--
The kinetic energy of two colliding particles evolve as
\begin{equation}\label{eq_evo_energy}
\begin{bmatrix}
  \tilde{\mathcal{E}}_i \\
  \tilde{\mathcal{E}}_{i+1}
\end{bmatrix}=\frac{1}{4}
\begin{bmatrix}
 \delta^2 & 4-\delta^2\\
 4-\delta^2 & \delta^2
\end{bmatrix}
\begin{bmatrix}
  \mathcal{E}_i \\
  \mathcal{E}_{i+1}
\end{bmatrix}+R
\begin{bmatrix}
  1 \\
  -1
\end{bmatrix},
\end{equation}
where $R={\rm sgn}(m_i-m_{i+1}) \delta\left(1-{\delta^2}/{4}\right)v_iv_{i+1}/2$. Equation (\ref{eq_evo_energy}) is a deterministic equation when $R=0$, i.e., $\delta=0$ or $\delta=2$, where the system is integrable. We define $\delta_2=2-\delta$ to measure the distance to the integrable point $\delta=2$.

According to the molecular chaos hypothesis (MCH, also known as \emph{Stosszahlansatz} \cite{boltzmann1964lectures,ehrenfest1990conceptual,Brown2009}), we have $\langle R\rangle=0$, and $\langle R^2\rangle =\mu\langle E_i E_{i+1}\rangle$, where $\mu=\delta^2\left(1-\delta^2/4\right)\in[0,1]$, which controls the randomness of Eq. (\ref{eq_evo_energy}). $\mu=0$ at $\delta=0$ or $\delta=2$. Following the MCH, the events of two particle collisions at any point in the system are statistically independent and equivalent. We then obtain a deterministic equation
\begin{equation}\label{eq-en}
\begin{bmatrix}
  \tilde{E}_i \\
  \tilde{E}_{i+1}
\end{bmatrix}=
\frac{1}{4}
\begin{bmatrix}
 \delta^2 & 4-\delta^2\\
 4-\delta^2 & \delta^2
\end{bmatrix}
\begin{bmatrix}
  E_i \\
  E_{i+1}
\end{bmatrix}\Leftrightarrow
\boldsymbol{\tilde{E}}=\boldsymbol{A}\boldsymbol{E},
\end{equation}
where $E=\langle \mathcal{E}\rangle$ is the ensemble average of the energy of per particle, i.e., Eq. (\ref{eq-en}) describes the evolution of the average energy over collisions, which is also a good approximation of Eq. (\ref{eq_evo_energy}) when $\mu$ is small.

The eigenvalues $\lambda$ and eigenvectors $\boldsymbol{u}$ of the matrix of coefficients $\boldsymbol{A}$ are, respectively, $\lambda_1=1$ and $ \boldsymbol{u}_1=[1,1]^{\rm T}$, which corresponds to the steady state of energy equipartition; and $\lambda_2={\delta^2}/{2}-1$, $\boldsymbol{u}_2=[-1,1]^{\rm T}$, which gives a unstable state. Because $|\lambda_2|<1$, which approaches zero after $n$ times collisions (i.e., $|\lambda_2|^n\to0$), when $\mu\neq0$ (i.e., nonintegrable case). In short, Eq. (\ref{eq-en}) has the unique stable fixed point, so the system described by Eq. (\ref{eq-en}) has the property of tending towards equilibrium when $\mu\neq0$. In addition, the rate $\chi$ of approaching equilibrium is given by the rate of $|\lambda_2|^n\to0$, i.e., $\chi\propto\delta^2$ for small $\delta$, and $\chi\propto\delta_2$ for small $\delta_2$. Besides, we note that $\delta=\sqrt{2}$ results in $\lambda_2=0$, Eq. (\ref{eq-en}) becomes $\tilde{E}_i=\tilde{E}_{i+1}=\frac{1}{2} \left(E_i+ E_{i+1}\right),$
which means that two particles reach equipartition after once collision. However, at the moment, $\mu=1$, i.e., the randomness of the system is the strongest, which can not be ignored; namely, Eq. (\ref{eq-en}) is not a good approximation of Eq. (\ref{eq_evo_energy}).

Following the above analysis, it is conjectured that near the integrable region the equipartition time $\mathcal{T}_{\rm eq}$ of the system is inversely proportional to the rate $\chi$. Moreover, consider the dependence of average collision time on temperature, i.e., Eq. (\ref{eq-meanTime}), we guess that
\begin{equation}\label{eq-Teq-delta}
  \mathcal{T}_{\rm eq}\propto
  \begin{cases}
    \delta^{-2} T^{-1/2}, & \delta\to0;\\
    \delta_2^{-1} T^{-1/2}, & \delta_2\to0,
  \end{cases}
\end{equation}
which will be verified numerically.

To observe the rate of thermalization, we define
\begin{equation}
  \Phi(t)=\frac{\sum_i E_i(t)^4}{(\sum_i E_i(t)^2)^2}\left[\frac{(\sum_i E_i(0)^2)^2}{\sum_i E_i(0)^4}\right],
\end{equation}
which is the inverse participation ratio (IPR) \cite{Wegner1980}. When $\delta=0$, the system is completely integrable, so the energy of the two colliding particles are exchanged with each other but not varied, namely, $\Phi(t)=1$ is unvaried. Only if $\delta\neq0$ does $\Phi$ evolve over time. It is expected that the ensemble average of $\langle\Phi(t)\rangle$ will approach a minimum value and no longer change over time when the system enters the thermalized state.

\begin{figure}[t]
  \centering
  \includegraphics[width=1\columnwidth]{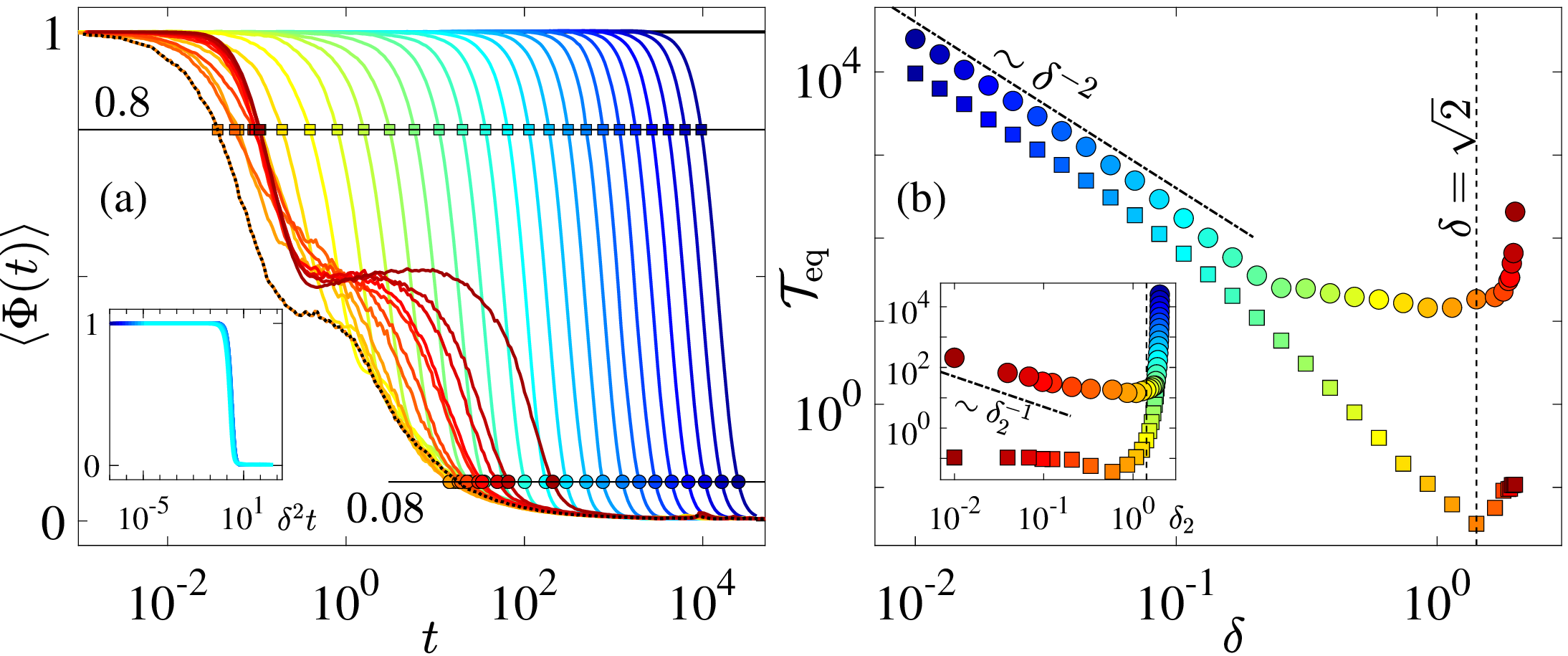}
  \caption{(a) The evolution of IPR $\langle\Phi(t)\rangle$ over time for various $\delta$. The horizontal lines of $0.8$ and $0.08$ are used to define the thermalization time $\mathcal{T}_{\rm eq}$, respectively. Inset: Same as the main panel but the horizontal coordinate rescaled by $\delta^2$, and $\delta \in[0.01,0.1]$. (b) $\mathcal{T}_{\rm eq}$ versus $\delta$. The data is directly taken from panel (a) and the color of the data points corresponds exactly to the color of the lines in panel (a). Inset: Same as the main panel but the horizontal coordinate is $\delta_2$. The dashed lines are plotted for reference. The system size $N=5000$ and the temperature $T=0.1$ are kept fixed.}\label{fig-Teq}
\end{figure}

\begin{figure}[t]
  \centering
  \includegraphics[width=1\columnwidth]{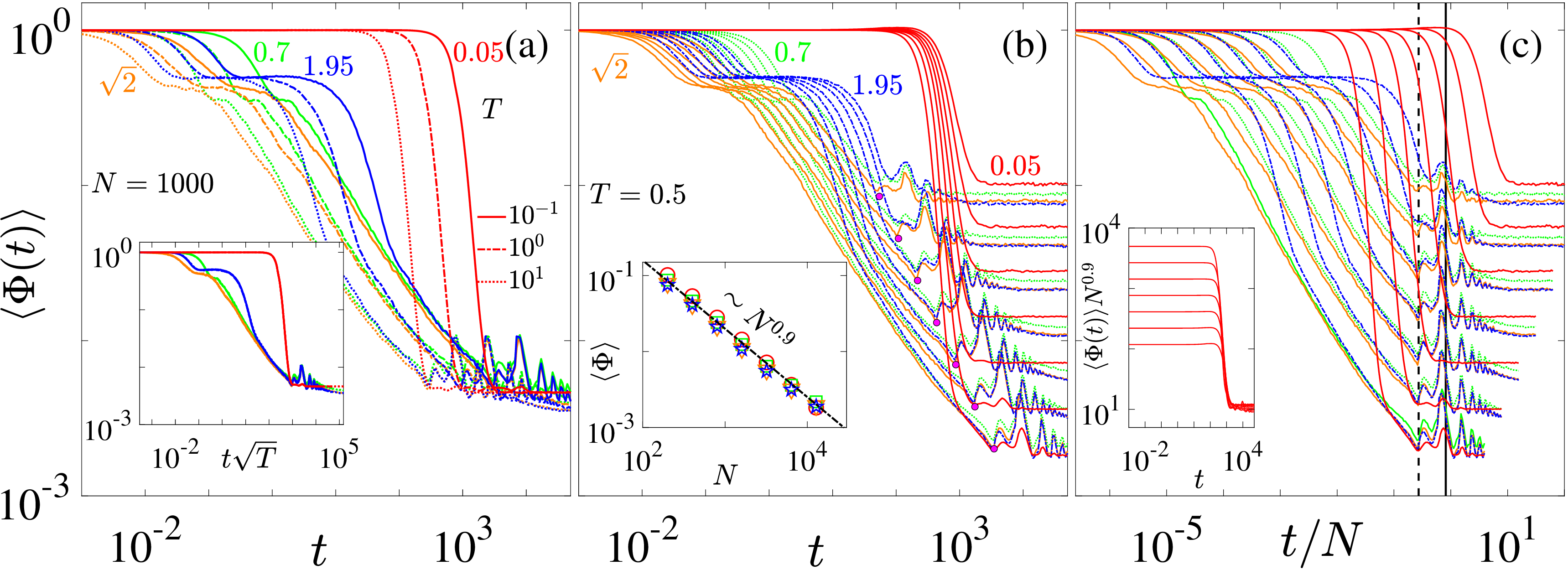}
  \caption{The evolution curve of IPR $\langle\Phi(t)\rangle$ over time.
  (a) The results for various temperatures. $N=5000$, and $\delta=0.1,0.7,\sqrt{2},1.95$ are kept fixed. Inset: Same as the main panel but the horizontal coordinate rescaled by $\sqrt{T}$. (b) Same as panel (a) but for different system sizes, see the red lines, from the top to the bottom, $N=100\times2^k$, $k=1, \dots, 8$, where $T=0.5$. Inset: The saturation value of $\langle\Phi\rangle$ as a function $N$. (c) Same as panel (b) but the horizontal coordinate rescaled by $1/N$. The two vertical lines correspond to $t/N=(3c_{\rm s})^{-1}$ and $t/N=c_{\rm s}^{-1}$, respectively. Inset: Same as panel (b) but only for the results of $\delta=0.05$, and $\langle\Phi\rangle$ is rescaled by $N^{0.9}$.}\label{fig-phi}
\end{figure}

Figure \ref{fig-Teq}(a) shows the numerical results of IPR $\langle \Phi(t)\rangle$ with various $\delta$. In our simulations, the system is evolved by applying an effective event-driven algorithm \cite{PhysRevE.67.015203}, and the periodic boundary conditions is adopted throughout. We see that $\langle \Phi(t)\rangle$ decreases to a stable value finally, and $\langle \Phi(t)\rangle$ can be renormalized to a line when $\delta$ is small (see the inset), while $\langle \Phi(t)\rangle$ has rich changing behavior when $\delta$ is large.

To clearly show the dependence of $\mathcal{T}_{\rm eq}$ on $\delta$, we define the time $\mathcal{T}(s,\delta)$ as a function of $s$ and $\delta$, where $\mathcal{T}$ is the time when $\langle \Phi(t)\rangle$ reaches the threshold value $s$, i.e., $\langle \Phi(\mathcal{T})\rangle=s$  for a given $\delta$. Figure \ref{fig-Teq}(b) presents $\mathcal{T}_{\rm eq}$ as a function $\delta$ for $s=0.8$ (squares) and $s=0.08$ (circles). We see that $\mathcal{T}_{\rm eq}$ is nonmonotonic, which is independent of $s$. For small $\delta$, $\mathcal{T}_{\rm eq}\propto\delta^{-2}$ (see main panel), which is consistent with the universal thermalization law observed recently in lattice systems \cite{PhysRevE.100.010101,Fu_2019,PhysRevE.100.052102,Onorato2019,PhysRevLett.124.186401,PhysRevE.104.L032104,Feng_2022}; while for larger $\delta$, $\mathcal{T}_{\rm eq}\propto\delta_2^{-1}$ at $s=0.08$ (see the inset). Besides, $\mathcal{T}_{\rm eq}$ achieves a minimum value at $\delta=\sqrt{2}$ in a short time (i.e., larger $s=0.8$) because $|A|=0$ when $\mu=1$, which means that the system reaches equipartition by once colliding if only considers the evolution of the deterministic part [see again Eq. (\ref{eq-en})]. However, due to the presence of random terms, this effect weakens over time, i.e., see circles for $s=0.08$.

Figure \ref{fig-phi}(a) shows the results of $\langle \Phi(t)\rangle$ at different temperatures. It can be seen that $\langle \Phi(t)\rangle$ has good scaling properties which is independent of $\delta$ (see the inset). It means that $\mathcal{T}_{\rm eq}\propto T^{-1/2}$ for all $\delta$. Figures \ref{fig-phi}(b) and \ref{fig-phi}(c) present the results of $\langle \Phi(t)\rangle$ at different sizes at fixed temperature. In the case of small $\delta$ (i.e., kinetic region), $\mathcal{T}_{\rm eq}$ is almost size-independent [see inset in Fig \ref{fig-phi}(c)], but in the case of large $\delta$ (i.e., hydrodynamics region), $\mathcal{T}_{\rm eq}$ has a strong size-dependence, where $\langle \Phi(t)\rangle$ attenuates to a certain degree and oscillates towards saturation, see the magenta points in Fig \ref{fig-phi}(b). If we define the time of thermalization as the time at which the oscillation begins, we have $\mathcal{T}_{\rm eq}\simeq \frac{N}{3c_{\rm s}\sqrt{T}}$ in the hydrodynamics region, see the vertical dashed line in Fig \ref{fig-phi}(c). According to the fluctuation dissipation theorem \cite{PhysRev.83.34}, there is a close relationship between the relaxation and transport properties of the system. Hereafter, we study the transport property of such a system.

\emph{Transport.}--
Similarly,
applying the MCH, the evolution of the local heat current is ruled by
\begin{equation}\label{eq-flux}
\begin{bmatrix}
  \tilde{\mathcal{J}}_i \\
  \tilde{\mathcal{J}}_{i+1}
\end{bmatrix}=
\begin{bmatrix}
 \left(\frac{m_i-m_{i+1}}{m_i+m_{i+1}}\right)^3 & \frac{8m_im_{i+1}^2}{(m_i+m_{i+1})^3}\\
 \frac{8m_i^2 m_{i+1}}{(m_i+m_{i+1})^3} & \left(\frac{m_{i+1}-m_i}{m_i+m_{i+1}}\right)^3
\end{bmatrix}
\begin{bmatrix}
  \mathcal{J}_i \\
  \mathcal{J}_{i+1}
\end{bmatrix},
\end{equation}
where $\mathcal{J}_i\equiv\frac{1}{2}m_iv_i^3$ and $\tilde{\mathcal{J}}_i\equiv\frac{1}{2}m_i\tilde{v}_i^3$ are the local heat current associated with the $i$th particle before and after a collision. Equation (\ref{eq-flux}) can be further abbreviated as $\boldsymbol{\mathcal{J}}_i(1)=\boldsymbol{W}_i\boldsymbol{\mathcal{J}}_i(0)$, where $\boldsymbol{W}_i$ is the matrix of coefficients, whose determinant is $|\boldsymbol{W}|=3\delta^2\left(1-{\delta^2}/{4}\right)/4-1\in[-1,-1/4]$, which is independent of the particle's label $i$. Notice that ${1}/{4}\leq||\boldsymbol{W}||\leq1$ which represents the scale factor by which the local areas (compressed) are transformed by $\boldsymbol{W}_i$ after a collision for $\mu\neq0$ since $||\boldsymbol{W}||<1$ \cite{LinearAlgebra2019}. While $||\boldsymbol{W}||=1$ when $\mu=0$, the areas does not change, which is the property of an integrable system.

Next we consider the heat current autocorrelation function (HCAF) of the system. For convenience, let us introduce a vector $\boldsymbol{J}=
\begin{bmatrix}
 \mathcal{J}_1, \mathcal{J}_2, \dots, \mathcal{J}_N
\end{bmatrix}^{\rm T}$. After $n$ times collisions, we have $\boldsymbol{J}(n)=\boldsymbol{\mathcal{W}}
\boldsymbol{J}(0)$, where $\boldsymbol{\mathcal{W}}=\prod_{{i_1},{i_2},\dots,{i_n}}^n
\boldsymbol{W}_{i_1}\boldsymbol{W}_{i_2}\dots\boldsymbol{W}_{i_n}$.
Hence, the HCAF can be written as
\begin{align}\label{eq-Cn}
  C(n)&=\langle\boldsymbol{J}^{\rm T}(n)\boldsymbol{J}(0)\rangle=\langle\boldsymbol{J}^{\rm T}(0)\boldsymbol{\mathcal{W}}^{\rm T}\boldsymbol{J}(0)\rangle\notag\\
  &=||\langle\boldsymbol{\mathcal{W}}^{\rm T}\rangle||\langle\boldsymbol{J}^{\rm T}(0)\boldsymbol{J}(0)\rangle\simeq C(0)||\boldsymbol{W}||^{\eta n},
\end{align}
where $\eta$ is a collision factor which needs to be determined (which is roughly estimated as $\eta\approx c_{\rm s}\theta/a\simeq\sqrt{3\pi}\simeq3$), and $\langle\cdot\rangle$ denotes ensemble average. Notice that here $n$ is the collision frequency rather than time. In general, the HCAF is a function of time, thus we rewrite Eq. (\ref{eq-Cn}) as
\begin{equation}\label{eq-Ct2}
  C(t)= C(0)||\boldsymbol{W}||^{\eta t/\theta}= C(0)e^{-t/\tau},
\end{equation}
where
\begin{equation}\label{eq-tau}
\tau = -\frac{\theta}{\eta\ln\left(||\boldsymbol{W}||\right)}
  =-\frac{\sqrt{{\pi(1-\delta^2/4)}/{T}}}{\eta\ln\left(1-\frac{3 \delta^2}{4}+\frac{3 \delta^4}{16}\right)},
\end{equation}
which is the characteristic time of exponential decay of HCAF. It is speculated that when $\tau\gg\theta$, the kinetic effects dominate the transport properties of the system, that is, the HCAF decreases exponentially; however, when $\tau$ and $\theta$ are approximately the same order of magnitude, or even $\tau\leq \theta$, the decay behavior of HCAF will change significantly, and the exponential decay region will almost disappear, which will be confirmed numerically latter.

Following the linear response theory, the thermal conductivity can be estimated, according to the Green-Kubo formula, as
\begin{equation}\label{eq-GK-kappa}
  \kappa = \lim_{t_{\rm c}\to\infty}\lim_{L\to\infty}\frac{1}{T^2L}\int_{0}^{t_{\rm c}}C(t)dt,
\end{equation}
where $t_{\rm c}=L/c_{\rm s}$ \cite{lepri2003thermal}. Insert Eq. (\ref{eq-Ct2}) into Eq. (\ref{eq-GK-kappa}), the heat conductivity contributed by the kinetic effect is
\begin{equation}\label{eq-kappa-cs}
  \kappa_{\rm k}(L)  =\lim_{t_{\rm c}\to\infty}\lim_{L\to\infty}\frac{C(0)~\tau}{T^2L}\left(1-e^{-\frac{t_{\rm c}}{\tau}}\right),
\end{equation}
where
\begin{equation}\label{eq-C0-PBC}
C(0) = \left\langle\left(\sum_{i=1}^{N}\mathcal{J}_i\right)^2\right\rangle
  =NT^3\left(\frac{15}{4-\delta^2}-\frac{9}{4}\right)
\end{equation}
under the periodic boundary conditions. When $L\gg c_{\rm s}\tau$, Eq. (\ref{eq-kappa-cs}) is simplified as
\begin{align}\label{eq-kappa-our0}
  \kappa_{\rm k} =T\left(\frac{15}{4-\delta^2}-\frac{9}{4}\right)\tau,
\end{align}
where the symbol for taking the limit has been omitted for brevity, which can be further concreted as
\begin{align}\label{eq-kappa-our}
  \kappa_{\rm k} =-\frac{1}{\eta}\left(\frac{15}{4-\delta^2}-\frac{9}{4}\right)
  \frac{\sqrt{\pi(1-\delta^2/4)T}}{\ln\left(1-\frac{3 \delta^2}{4}+\frac{3 \delta^4}{16}\right)},
\end{align}
where $\eta$ is the only parameter that cannot be settled by the initial conditions. The literature \cite{PhysRevE.90.032134} shows that the ratio of thermal conductivity at different temperatures is ${\kappa(T')}/{\kappa(T)}={\sqrt{T'}}/{\sqrt{T}}$, which means that $\eta$ should be independent of the temperature if Eq. (\ref{eq-kappa-our}) holds. Notice that $\eta$ can only be a function of $N$ and $\delta$. Next, we will ascertain parameter $\eta$ through numerical experiments.

\begin{figure}[t]
  \centering
  \includegraphics[width=1\columnwidth]{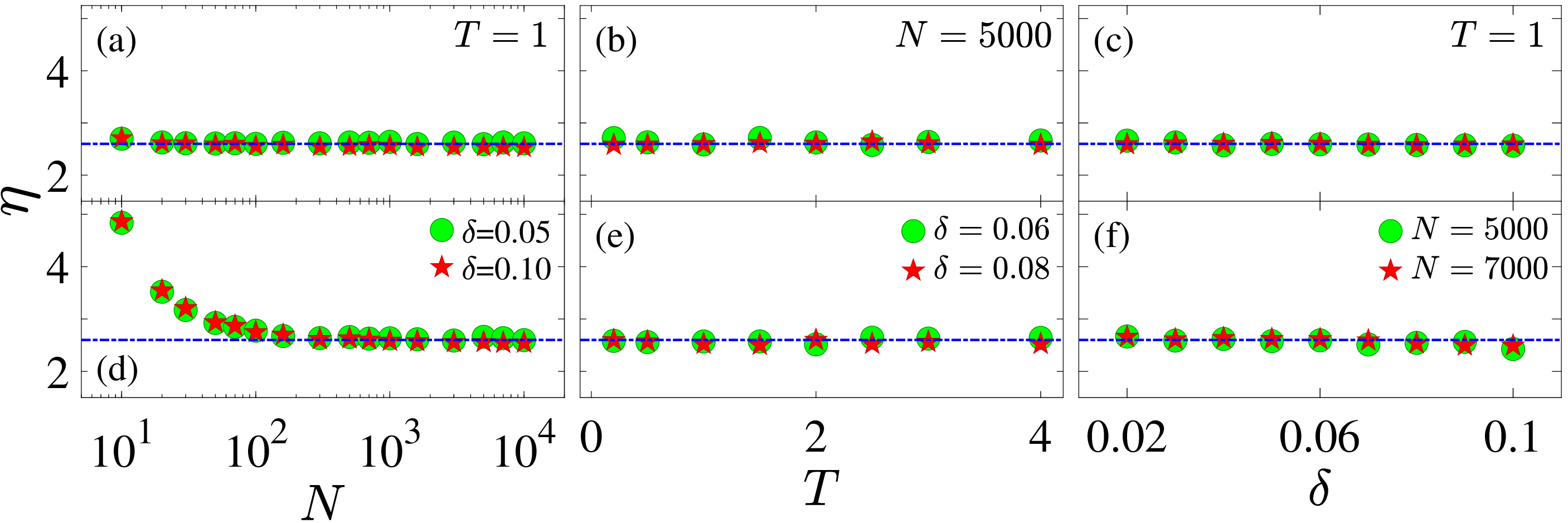}
  \caption{The dependence of the collision factor $\eta$ on the system size $N$ (a)/(d), temperature $T$ (b)/(e), and mass difference $\delta$ (c)/(f). The data in panels (a)-(c) are obtained by fitting the decay exponent of the HCAF, i.e., Eqs. (\ref{eq-Ct2}) and (\ref{eq-tau}). The data in panels (d)-(f) are acquired through Eq. (\ref{eq-kappa-our}). The horizontal blue doted-dashed lines in all panels correspond to $\eta=2.6$, which are plotted for reference.}\label{fig-eta}
\end{figure}

\begin{figure}[t]
  \centering
  \includegraphics[width=1\columnwidth]{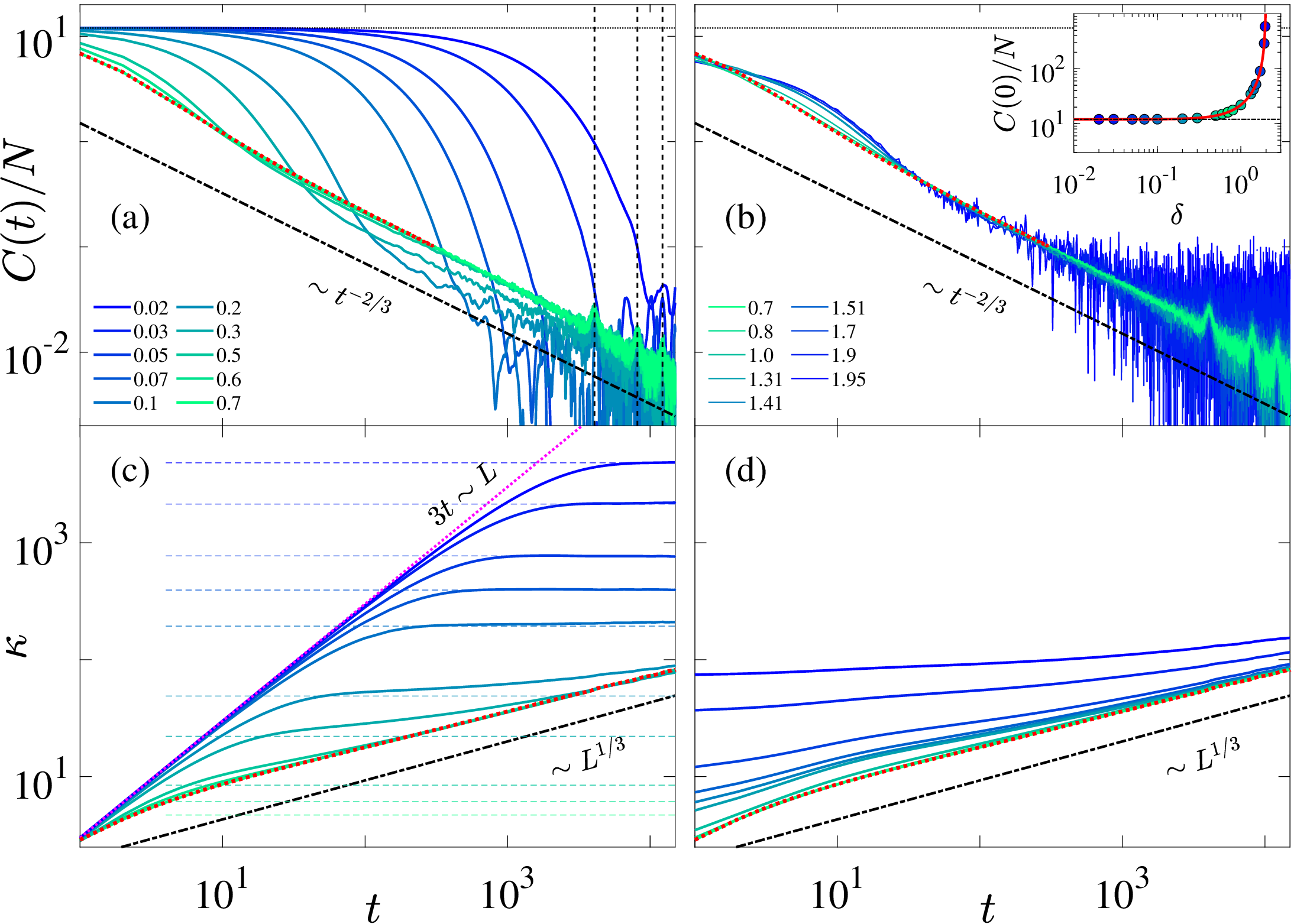}
  \caption{(a) and (b) are the normalized HCAF, i.e.,$C(t)/N$, for various $\delta$. The horizontal lines in (a) and (b) are the value of $C(0)/N$ with $\delta=0$ for reference, see Eq. (\ref{eq-C0-PBC}). The vertical dashed lines are for $t=4073$, $8146$, and $12219$. Inset in panel (b) shows $C(0)/N$ as a function of $\delta$. The red solid line is the prediction of Eq. (\ref{eq-C0-PBC}). (c) and (d) are heat conductivity calculated from (a) and (b) through Eq. (\ref{eq-GK-kappa}). The horizontal dashed lines in panel (c) are theoretical predictions, see Eq. (\ref{eq-kappa-our}). The magenta dotted line is $\kappa=3t$ given by Eq. (\ref{eq-kappa-our0}) for $\delta=0$. In all panels: the black dashed-dotted lines with different slopes are drawn for reference, and the red dotted lines cover the result of $\delta=0.7$ for easy identifying. The system size $N=10^4$ and the temperature $T=2$ are fixed.}\label{fig-corr-kappa}
\end{figure}

Figure \ref{fig-eta} shows the collision factor $\eta$ as a function of the system size $N$, temperature $T$, and mass difference $\delta$, respectively. Here $\eta$ is obtained in two ways. In the first method, we calculate the HCAF through molecular dynamics simulation under given parameters, and then directly fit the characteristic time $\tau$ (see the exponential decay of HCAF in Fig. \ref{fig-corr-kappa}(a)) to obtain $\eta$ according to Eq. (\ref{eq-tau}). The results are presented in Figs. \ref{fig-eta}(a)-(c). In the second method, we integrate the HCAF numerically. The integral value will saturate to a stable value for small $\delta$ (see the plateau in Fig. \ref{fig-corr-kappa}(c)), and then $\eta$ is obtained based on expression (\ref{eq-kappa-our}). The results are plotted in Figs. \ref{fig-eta}(d)-(f). It is shown that $\eta\simeq2.6$ is a constant independent of $N$, $T$, and $\delta$. Although the results in Fig. \ref{fig-eta} (d) show that $\eta$ increases at small $N$, this is because we use expression (\ref{eq-kappa-our}) to calculate. In principle, in the case of small sizes, Eq. (\ref{eq-kappa-cs}) should be used for calculation, but solving the transcendental equation is more complicated, so we use Eq. (\ref{eq-kappa-our}) for calculation, which results in the deviations. Since $\eta$ is a constant, for small $\delta$, we have $\tau \propto \delta^{-2}T^{-1/2}$, and $\kappa_{\rm k}\propto \delta^{-2}T^{1/2}$; while for small $\delta_2$, we have $\tau \propto \delta_2^{-1}T^{-1/2}$, and $\kappa_{\rm k}\propto \delta_2^{-1}T^{1/2}$.

\begin{figure}[t]
  \centering
  \includegraphics[width=1\columnwidth]{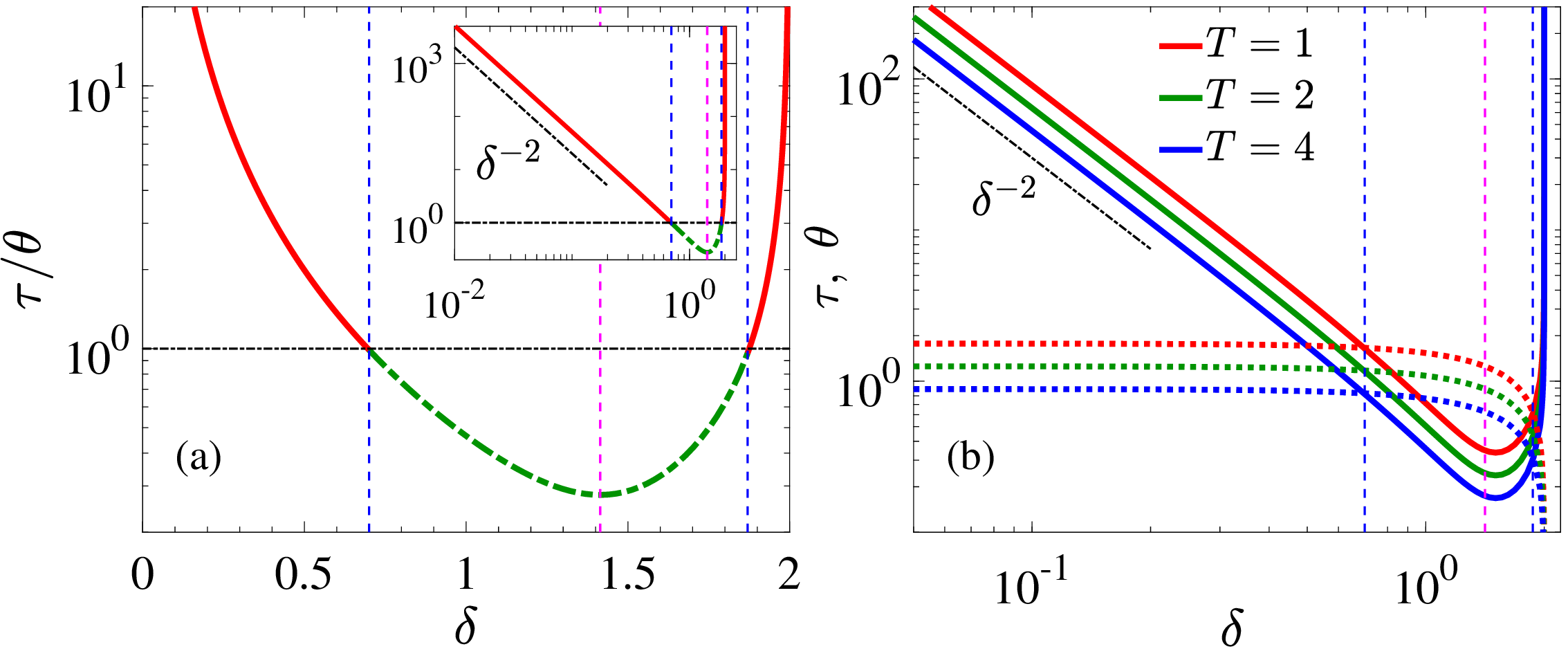}
  \caption{(a) The ratio $\tau/\theta$ as a function of $\delta$. The green dotted line is for $\tau/\theta<1$. Inset: Same as the main panel but in log-log scale. (b) Dependence of $\tau$ and $\theta$ on $\delta$ with different $T$. The vertical dashed lines in all panels are, respectively, for $\delta=0.7$, $\sqrt{2}$, and $1.87$, which are plotted for reference.}\label{fig-tau}
\end{figure}

In Fig. \ref{fig-tau}(a), we show the dependence of the ratio $\tau/\theta$ on $\delta$, which is independent of the temperature. We see that $\tau/\theta$ is nonmonotonic and reaches its minimum at $\delta=\sqrt{2}$, while diverging at $\delta=0$ and $2$. Besides, $\tau/\theta\approx1$ at $\delta=0.7$ and $\delta=1.87$. We plot the function curves of $\tau$ and $\theta$ with different temperatures in Fig. \ref{fig-tau}(b). It is seen that $\tau$ and $\theta$ intersect at $\delta=0.7$ and $\delta=1.87$ at different temperatures. Next, we will study the behavior of HCAF in the systems with different $\delta$, focusing on the behavior around $\delta=0.7$.

In Figs. \ref{fig-corr-kappa}(a) and \ref{fig-corr-kappa}(b), we show the evolution of $C(t)/N$ for various $\delta$. As shown in Fig. \ref{fig-corr-kappa}(a), the HCAF decreases exponentially for small $\delta$. As $\delta$ gradually increases, the region of exponential decay decreases, especially when $\delta=0.7$, the region of exponential decay almost completely disappears (see red dotted), and the whole decays in a manner of $C(t)\sim t^{-2/3}$, which is a standard hydrodynamics behavior. Besides, the oscillations in HCAF are clearly observed [see the vertical dashed lines at $t=4073$, thus $c_{\rm s}=10^4/4073\approx2.455$, which agrees with $c_{\rm s}=\sqrt{6}\approx 2.449$, see again Eq. (\ref{eq-cs})], which is caused by the recurrence \cite{PhysRevE.89.022111}. From Fig. \ref{fig-corr-kappa}(b), we see that the HCAF remains power-law decay over a large range as $\delta$ continues to increase, but a small region (see $2<t<30$) deviates from the power-law for larger $\delta$. Inset in panel \ref{fig-corr-kappa}(b) presents $C(0)/N$ as a function of $\delta$. $C(0)/N$ monotonically increases with the increase of $\delta$, and when $\delta$ approaches $2$, $C(0)/N$ diverges.

Figures \ref{fig-corr-kappa}(c) and \ref{fig-corr-kappa}(d) show the heat conductivity $\kappa$ as a function of time $t$, which are, respectively, calculated from \ref{fig-corr-kappa}(a) and \ref{fig-corr-kappa}(b) through Eq. (\ref{eq-GK-kappa}). From Fig. \ref{fig-corr-kappa}(c), we see that when $\delta$ is small, $\kappa$ saturates a plateau, which is described by expression (\ref{eq-kappa-our}). However, with the increase of $\delta$, $\kappa$ gradually appears to increase, and eventually tends to the behavior $\kappa\propto L^{1/3}$ predicted by the hydrodynamics theory \cite{PhysRevLett.89.200601,PhysRevLett.89.180601,PhysRevLett.108.180601,PhysRevLett.111.230601}.
Intriguingly, we see that the $\kappa$ is minimal at $\delta=0.7$ for a given size since $\tau/\theta\simeq1$ results in $\kappa_{\rm k}$ to be minimum. In addition, for $\delta=0$, according to Eq. (\ref{eq-kappa-our0}), $\kappa_{\rm k}=\frac{3T}{2}\frac{L}{c_{\rm s}}=3t$, see the magenta dotted line, which agrees well with the predictions of the Debye's theory for $\delta=0$, i.e., $\kappa=\mathcal{C} L c_{\rm s} =\frac{L}{2}\sqrt{3T}=\kappa_{\rm k}$, where $\mathcal{C}=1/2$ is the heat capacity of 1D gases \cite{lepri2003thermal}. In fact, we know that the HCAF is a constant that does not change with time. However, Eq. (\ref{eq-kappa-our0}) is derived from the exponential decay of the HCAF. The results in Fig. \ref{fig-corr-kappa}(d) show that with the further increase of $\delta$, the value of $\kappa(t=1)$ becomes larger and larger since the larger $\delta$, the larger $C(0)$ [see again inset in panel \ref{fig-corr-kappa}(b)], and the whole curve tends to line of $\kappa\propto L^{1/3}$ at a very slow speed. Because an increase in $\delta$ leads to the system tending towards another integrable limit ($\delta\simeq2$), therefore, it deviates from the prediction for larger $\delta$. Within the near integrable region, it is expected that a much larger $L$ will be needed to observe $\kappa\propto L^{1/3}$. Following we will make a rough estimate of the size required.

Assuming that the HCAF decays exponentially first and then changes into a power-law way decay at $\tau_c$, see again Fig. \ref{fig-corr-kappa}(a). The thermal conductivity contributed by the hydrodynamics effects can be estimated, according to Eq. (\ref{eq-GK-kappa}), as
\begin{equation}\label{eq-kappa-hydro}
\kappa_{\rm h}(L)=
\frac{3C(0)\tau_c}{T^{2}L}e^{-\frac{\tau_c}{\tau}}
\left[\left(\frac{L}{c_{\rm s}\tau_c}\right)^{1/3}  -1\right],
\end{equation}
where $L>c_{\rm s}\tau_c$. In the near integrable region, $\tau_c$ is at least several times of $\tau$, but we assume that $\tau_c=\tau$, and we set $\lambda={\kappa_{\rm h}}/{\kappa_{\rm k}}$, then we have $L =
c_{\rm s}\tau\left[{\lambda\left(e-1\right)}/{3}+1\right]^3\propto \delta^{-2}$, which means that $L\to\infty$ when $\delta\to0$, namely, the critical size $L_{\rm c}$ required for growth on the platform in Fig. \ref{fig-corr-kappa}(c) is divergent when $\delta\to0$. This may be the reason for seeing different divergence exponents in 1D systems \cite{PhysRevE.85.020102,pnas2015,PhysRevE.97.022116}.

\emph{Summary.}--We have studied the thermalization and transport properties of a 1D DHP system, which has two integrable reference points. We show that $\mathcal{T}_{\rm eq}\propto \delta^{-2} T^{-1/2}$ for $\delta\to0$, which agree with the universal thermalization behavior of the lattices in the vicinity of integrable limit; and $\mathcal{T}_{\rm eq}\propto (2-\delta)^{-1} T^{-1/2}$ for $\delta\to2$, which is corresponding to the tendency of independent oscillators to thermalization. In particular, the system exhibits normal heat conduction behavior when $\delta\to0$ (i.e., kinetics is dominant), since only if $L>L_{\rm c}\propto\delta^{-2}$ to observe $\kappa\sim L^{1/3}$. Namely, the hydrodynamic behavior can not be observed when $\delta\to0$. However, the behavior of the system is dominated by hydrodynamics when $0.7<\delta<1.87$, where $\mathcal{T}_{\rm eq}\simeq \frac{N}{3c_{\rm s}\sqrt{T}}$ and $\kappa\sim N^{1/3}$, that is, the relaxation time and transport coefficient of the system are size-dependent. The size-independent or weakly size-dependent transport coefficient of the system in the kinetic dominant region has important guiding significance for us to search for thermoelectric materials with high thermoelectric figure of merit \cite{PhysRevLett.110.070604,Chen2015,PhysRevLett.121.080602}.

Furthermore, our findings suggest that the approaches employed in this study can be extended to investigate related problems in perturbed Toda lattices \cite{Fu_2019,PhysRevE.100.052102,PhysRevE.85.060102,Benettin2023}. This opens up a plethora of opportunities for future research, where the interplay of various factors in more complex systems can be explored.

\section*{Acknowledgment}
We acknowledge support by the NSFC (Grants No. 12005156, No. 11975190, No. 12247106, and No. 12247101), and by the Natural Science Foundation of Gansu Province (Grants No. 21JR1RE289, and No. 20JR5RA494), and by the Innovation Fund from Department of Education of Gansu Province (Grant No. 2023A-106), and by the Project of Fu-Xi Scientific Research Innovation Team, Tianshui Normal University (Grant No. FXD2020-02), and by the Education Project of Open Competition for the Best Candidates from Department of Education of Gansu Province, China (Grant No. 2021jyjbgs-06).

\bibliography{refgas}

\begin{thebibliography}{40}%
\makeatletter
\providecommand \@ifxundefined [1]{%
 \@ifx{#1\undefined}
}%
\providecommand \@ifnum [1]{%
 \ifnum #1\expandafter \@firstoftwo
 \else \expandafter \@secondoftwo
 \fi
}%
\providecommand \@ifx [1]{%
 \ifx #1\expandafter \@firstoftwo
 \else \expandafter \@secondoftwo
 \fi
}%
\providecommand \natexlab [1]{#1}%
\providecommand \enquote  [1]{``#1''}%
\providecommand \bibnamefont  [1]{#1}%
\providecommand \bibfnamefont [1]{#1}%
\providecommand \citenamefont [1]{#1}%
\providecommand \href@noop [0]{\@secondoftwo}%
\providecommand \href [0]{\begingroup \@sanitize@url \@href}%
\providecommand \@href[1]{\@@startlink{#1}\@@href}%
\providecommand \@@href[1]{\endgroup#1\@@endlink}%
\providecommand \@sanitize@url [0]{\catcode `\\12\catcode `\$12\catcode
  `\&12\catcode `\#12\catcode `\^12\catcode `\_12\catcode `\%12\relax}%
\providecommand \@@startlink[1]{}%
\providecommand \@@endlink[0]{}%
\providecommand \url  [0]{\begingroup\@sanitize@url \@url }%
\providecommand \@url [1]{\endgroup\@href {#1}{\urlprefix }}%
\providecommand \urlprefix  [0]{URL }%
\providecommand \Eprint [0]{\href }%
\providecommand \doibase [0]{http://dx.doi.org/}%
\providecommand \selectlanguage [0]{\@gobble}%
\providecommand \bibinfo  [0]{\@secondoftwo}%
\providecommand \bibfield  [0]{\@secondoftwo}%
\providecommand \translation [1]{[#1]}%
\providecommand \BibitemOpen [0]{}%
\providecommand \bibitemStop [0]{}%
\providecommand \bibitemNoStop [0]{.\EOS\space}%
\providecommand \EOS [0]{\spacefactor3000\relax}%
\providecommand \BibitemShut  [1]{\csname bibitem#1\endcsname}%
\let\auto@bib@innerbib\@empty
%</preamble>
\bibitem [{\citenamefont {Casati}\ and\ \citenamefont
  {Ford}(1976)}]{casati1976computer}%
  \BibitemOpen
  \bibfield  {author} {\bibinfo {author} {\bibfnamefont {G.}~\bibnamefont
  {Casati}}\ and\ \bibinfo {author} {\bibfnamefont {J.}~\bibnamefont {Ford}},\
  }\href {\doibase https://doi.org/10.1016/0021-9991(76)90104-2} {\bibfield
  {journal} {\bibinfo  {journal} {J. Comput. Phys.}\ }\textbf {\bibinfo
  {volume} {20}},\ \bibinfo {pages} {97} (\bibinfo {year} {1976})}\BibitemShut
  {NoStop}%
\bibitem [{\citenamefont {Casati}(1986)}]{casati1986energy}%
  \BibitemOpen
  \bibfield  {author} {\bibinfo {author} {\bibfnamefont {G.}~\bibnamefont
  {Casati}},\ }\href {\doibase 10.1007/BF00735180} {\bibfield  {journal}
  {\bibinfo  {journal} {Found. Phys.}\ }\textbf {\bibinfo {volume} {16}},\
  \bibinfo {pages} {51} (\bibinfo {year} {1986})}\BibitemShut {NoStop}%
\bibitem [{\citenamefont {Casati}\ and\ \citenamefont
  {Prosen}(2003)}]{PhysRevE.67.015203}%
  \BibitemOpen
  \bibfield  {author} {\bibinfo {author} {\bibfnamefont {G.}~\bibnamefont
  {Casati}}\ and\ \bibinfo {author} {\bibfnamefont {T.}~\bibnamefont
  {Prosen}},\ }\href {\doibase 10.1103/PhysRevE.67.015203} {\bibfield
  {journal} {\bibinfo  {journal} {Phys. Rev. E}\ }\textbf {\bibinfo {volume}
  {67}},\ \bibinfo {pages} {015203} (\bibinfo {year} {2003})}\BibitemShut
  {NoStop}%
\bibitem [{\citenamefont {Chen}\ \emph
  {et~al.}(2014{\natexlab{a}})\citenamefont {Chen}, \citenamefont {Zhang},
  \citenamefont {Wang},\ and\ \citenamefont {Zhao}}]{PhysRevE.89.022111}%
  \BibitemOpen
  \bibfield  {author} {\bibinfo {author} {\bibfnamefont {S.}~\bibnamefont
  {Chen}}, \bibinfo {author} {\bibfnamefont {Y.}~\bibnamefont {Zhang}},
  \bibinfo {author} {\bibfnamefont {J.}~\bibnamefont {Wang}}, \ and\ \bibinfo
  {author} {\bibfnamefont {H.}~\bibnamefont {Zhao}},\ }\href {\doibase
  10.1103/PhysRevE.89.022111} {\bibfield  {journal} {\bibinfo  {journal} {Phys.
  Rev. E}\ }\textbf {\bibinfo {volume} {89}},\ \bibinfo {pages} {022111}
  (\bibinfo {year} {2014}{\natexlab{a}})}\BibitemShut {NoStop}%
\bibitem [{\citenamefont {Li}\ \emph {et~al.}(2004)\citenamefont {Li},
  \citenamefont {Casati}, \citenamefont {Wang},\ and\ \citenamefont
  {Prosen}}]{Li2004}%
  \BibitemOpen
  \bibfield  {author} {\bibinfo {author} {\bibfnamefont {B.}~\bibnamefont
  {Li}}, \bibinfo {author} {\bibfnamefont {G.}~\bibnamefont {Casati}}, \bibinfo
  {author} {\bibfnamefont {J.}~\bibnamefont {Wang}}, \ and\ \bibinfo {author}
  {\bibfnamefont {T.~c.~v.}\ \bibnamefont {Prosen}},\ }\href {\doibase
  10.1103/PhysRevLett.92.254301} {\bibfield  {journal} {\bibinfo  {journal}
  {Phys. Rev. Lett.}\ }\textbf {\bibinfo {volume} {92}},\ \bibinfo {pages}
  {254301} (\bibinfo {year} {2004})}\BibitemShut {NoStop}%
\bibitem [{\citenamefont {Benenti}\ \emph {et~al.}(2013)\citenamefont
  {Benenti}, \citenamefont {Casati},\ and\ \citenamefont
  {Wang}}]{PhysRevLett.110.070604}%
  \BibitemOpen
  \bibfield  {author} {\bibinfo {author} {\bibfnamefont {G.}~\bibnamefont
  {Benenti}}, \bibinfo {author} {\bibfnamefont {G.}~\bibnamefont {Casati}}, \
  and\ \bibinfo {author} {\bibfnamefont {J.}~\bibnamefont {Wang}},\ }\href
  {\doibase 10.1103/PhysRevLett.110.070604} {\bibfield  {journal} {\bibinfo
  {journal} {Phys. Rev. Lett.}\ }\textbf {\bibinfo {volume} {110}},\ \bibinfo
  {pages} {070604} (\bibinfo {year} {2013})}\BibitemShut {NoStop}%
\bibitem [{\citenamefont {Luo}\ \emph {et~al.}(2018)\citenamefont {Luo},
  \citenamefont {Benenti}, \citenamefont {Casati},\ and\ \citenamefont
  {Wang}}]{PhysRevLett.121.080602}%
  \BibitemOpen
  \bibfield  {author} {\bibinfo {author} {\bibfnamefont {R.}~\bibnamefont
  {Luo}}, \bibinfo {author} {\bibfnamefont {G.}~\bibnamefont {Benenti}},
  \bibinfo {author} {\bibfnamefont {G.}~\bibnamefont {Casati}}, \ and\ \bibinfo
  {author} {\bibfnamefont {J.}~\bibnamefont {Wang}},\ }\href {\doibase
  10.1103/PhysRevLett.121.080602} {\bibfield  {journal} {\bibinfo  {journal}
  {Phys. Rev. Lett.}\ }\textbf {\bibinfo {volume} {121}},\ \bibinfo {pages}
  {080602} (\bibinfo {year} {2018})}\BibitemShut {NoStop}%
\bibitem [{\citenamefont {Dhar}(2001)}]{PhysRevLett.86.3554}%
  \BibitemOpen
  \bibfield  {author} {\bibinfo {author} {\bibfnamefont {A.}~\bibnamefont
  {Dhar}},\ }\href {\doibase 10.1103/PhysRevLett.86.3554} {\bibfield  {journal}
  {\bibinfo  {journal} {Phys. Rev. Lett.}\ }\textbf {\bibinfo {volume} {86}},\
  \bibinfo {pages} {3554} (\bibinfo {year} {2001})}\BibitemShut {NoStop}%
\bibitem [{\citenamefont {Boozer}(2011)}]{PhysRevE.84.031127}%
  \BibitemOpen
  \bibfield  {author} {\bibinfo {author} {\bibfnamefont {A.~D.}\ \bibnamefont
  {Boozer}},\ }\href {\doibase 10.1103/PhysRevE.84.031127} {\bibfield
  {journal} {\bibinfo  {journal} {Phys. Rev. E}\ }\textbf {\bibinfo {volume}
  {84}},\ \bibinfo {pages} {031127} (\bibinfo {year} {2011})}\BibitemShut
  {NoStop}%
\bibitem [{\citenamefont {Grassberger}\ \emph {et~al.}(2002)\citenamefont
  {Grassberger}, \citenamefont {Nadler},\ and\ \citenamefont
  {Yang}}]{PhysRevLett.89.180601}%
  \BibitemOpen
  \bibfield  {author} {\bibinfo {author} {\bibfnamefont {P.}~\bibnamefont
  {Grassberger}}, \bibinfo {author} {\bibfnamefont {W.}~\bibnamefont {Nadler}},
  \ and\ \bibinfo {author} {\bibfnamefont {L.}~\bibnamefont {Yang}},\ }\href
  {\doibase 10.1103/PhysRevLett.89.180601} {\bibfield  {journal} {\bibinfo
  {journal} {Phys. Rev. Lett.}\ }\textbf {\bibinfo {volume} {89}},\ \bibinfo
  {pages} {180601} (\bibinfo {year} {2002})}\BibitemShut {NoStop}%
\bibitem [{\citenamefont {Cipriani}\ \emph {et~al.}(2005)\citenamefont
  {Cipriani}, \citenamefont {Denisov},\ and\ \citenamefont
  {Politi}}]{PhysRevLett.94.244301}%
  \BibitemOpen
  \bibfield  {author} {\bibinfo {author} {\bibfnamefont {P.}~\bibnamefont
  {Cipriani}}, \bibinfo {author} {\bibfnamefont {S.}~\bibnamefont {Denisov}}, \
  and\ \bibinfo {author} {\bibfnamefont {A.}~\bibnamefont {Politi}},\ }\href
  {\doibase 10.1103/PhysRevLett.94.244301} {\bibfield  {journal} {\bibinfo
  {journal} {Phys. Rev. Lett.}\ }\textbf {\bibinfo {volume} {94}},\ \bibinfo
  {pages} {244301} (\bibinfo {year} {2005})}\BibitemShut {NoStop}%
\bibitem [{\citenamefont {Lepri}\ \emph {et~al.}(2003)\citenamefont {Lepri},
  \citenamefont {Livi},\ and\ \citenamefont {Politi}}]{lepri2003thermal}%
  \BibitemOpen
  \bibfield  {author} {\bibinfo {author} {\bibfnamefont {S.}~\bibnamefont
  {Lepri}}, \bibinfo {author} {\bibfnamefont {R.}~\bibnamefont {Livi}}, \ and\
  \bibinfo {author} {\bibfnamefont {A.}~\bibnamefont {Politi}},\ }\href
  {\doibase 10.1016/S0370-1573(02)00558-6} {\bibfield  {journal} {\bibinfo
  {journal} {Phys. Rep.}\ }\textbf {\bibinfo {volume} {377}},\ \bibinfo {pages}
  {1} (\bibinfo {year} {2003})}\BibitemShut {NoStop}%
\bibitem [{\citenamefont {Dhar}(2008)}]{Dhar08AdvPhys}%
  \BibitemOpen
  \bibfield  {author} {\bibinfo {author} {\bibfnamefont {A.}~\bibnamefont
  {Dhar}},\ }\href {\doibase 10.1080/00018730802538522} {\bibfield  {journal}
  {\bibinfo  {journal} {Adv Phys}\ }\textbf {\bibinfo {volume} {57}},\ \bibinfo
  {pages} {457} (\bibinfo {year} {2008})}\BibitemShut {NoStop}%
\bibitem [{\citenamefont {Narayan}\ and\ \citenamefont
  {Ramaswamy}(2002)}]{PhysRevLett.89.200601}%
  \BibitemOpen
  \bibfield  {author} {\bibinfo {author} {\bibfnamefont {O.}~\bibnamefont
  {Narayan}}\ and\ \bibinfo {author} {\bibfnamefont {S.}~\bibnamefont
  {Ramaswamy}},\ }\href {\doibase 10.1103/PhysRevLett.89.200601} {\bibfield
  {journal} {\bibinfo  {journal} {Phys. Rev. Lett.}\ }\textbf {\bibinfo
  {volume} {89}},\ \bibinfo {pages} {200601} (\bibinfo {year}
  {2002})}\BibitemShut {NoStop}%
\bibitem [{\citenamefont {van Beijeren}(2012)}]{PhysRevLett.108.180601}%
  \BibitemOpen
  \bibfield  {author} {\bibinfo {author} {\bibfnamefont {H.}~\bibnamefont {van
  Beijeren}},\ }\href {\doibase 10.1103/PhysRevLett.108.180601} {\bibfield
  {journal} {\bibinfo  {journal} {Phys. Rev. Lett.}\ }\textbf {\bibinfo
  {volume} {108}},\ \bibinfo {pages} {180601} (\bibinfo {year}
  {2012})}\BibitemShut {NoStop}%
\bibitem [{\citenamefont {Mendl}\ and\ \citenamefont
  {Spohn}(2013)}]{PhysRevLett.111.230601}%
  \BibitemOpen
  \bibfield  {author} {\bibinfo {author} {\bibfnamefont {C.~B.}\ \bibnamefont
  {Mendl}}\ and\ \bibinfo {author} {\bibfnamefont {H.}~\bibnamefont {Spohn}},\
  }\href {\doibase 10.1103/PhysRevLett.111.230601} {\bibfield  {journal}
  {\bibinfo  {journal} {Phys. Rev. Lett.}\ }\textbf {\bibinfo {volume} {111}},\
  \bibinfo {pages} {230601} (\bibinfo {year} {2013})}\BibitemShut {NoStop}%
\bibitem [{\citenamefont {Chen}\ \emph
  {et~al.}(2014{\natexlab{b}})\citenamefont {Chen}, \citenamefont {Wang},
  \citenamefont {Casati},\ and\ \citenamefont {Benenti}}]{PhysRevE.90.032134}%
  \BibitemOpen
  \bibfield  {author} {\bibinfo {author} {\bibfnamefont {S.}~\bibnamefont
  {Chen}}, \bibinfo {author} {\bibfnamefont {J.}~\bibnamefont {Wang}}, \bibinfo
  {author} {\bibfnamefont {G.}~\bibnamefont {Casati}}, \ and\ \bibinfo {author}
  {\bibfnamefont {G.}~\bibnamefont {Benenti}},\ }\href {\doibase
  10.1103/PhysRevE.90.032134} {\bibfield  {journal} {\bibinfo  {journal} {Phys.
  Rev. E}\ }\textbf {\bibinfo {volume} {90}},\ \bibinfo {pages} {032134}
  (\bibinfo {year} {2014}{\natexlab{b}})}\BibitemShut {NoStop}%
\bibitem [{\citenamefont {Zhao}\ and\ \citenamefont
  {Wang}(2018)}]{PhysRevE.97.010103}%
  \BibitemOpen
  \bibfield  {author} {\bibinfo {author} {\bibfnamefont {H.}~\bibnamefont
  {Zhao}}\ and\ \bibinfo {author} {\bibfnamefont {W.-g.}\ \bibnamefont
  {Wang}},\ }\href {\doibase 10.1103/PhysRevE.97.010103} {\bibfield  {journal}
  {\bibinfo  {journal} {Phys. Rev. E}\ }\textbf {\bibinfo {volume} {97}},\
  \bibinfo {pages} {010103} (\bibinfo {year} {2018})}\BibitemShut {NoStop}%
\bibitem [{\citenamefont {Lepri}\ \emph {et~al.}(2020)\citenamefont {Lepri},
  \citenamefont {Livi},\ and\ \citenamefont {Politi}}]{PhysRevLett.125.040604}%
  \BibitemOpen
  \bibfield  {author} {\bibinfo {author} {\bibfnamefont {S.}~\bibnamefont
  {Lepri}}, \bibinfo {author} {\bibfnamefont {R.}~\bibnamefont {Livi}}, \ and\
  \bibinfo {author} {\bibfnamefont {A.}~\bibnamefont {Politi}},\ }\href
  {\doibase 10.1103/PhysRevLett.125.040604} {\bibfield  {journal} {\bibinfo
  {journal} {Phys. Rev. Lett.}\ }\textbf {\bibinfo {volume} {125}},\ \bibinfo
  {pages} {040604} (\bibinfo {year} {2020})}\BibitemShut {NoStop}%
\bibitem [{\citenamefont {Zuckerwar}(2002)}]{wong2002handbook}%
  \BibitemOpen
  \bibfield  {author} {\bibinfo {author} {\bibfnamefont {A.~J.}\ \bibnamefont
  {Zuckerwar}},\ }\href@noop {} {\emph {\bibinfo {title} {Handbook of the Speed
  of Sound in Real Gases}}}\ (\bibinfo  {publisher} {Academic Press},\ \bibinfo
  {year} {2002})\BibitemShut {NoStop}%
\bibitem [{\citenamefont {Bogoliubov}(1962)}]{bogoliubov1962problems}%
  \BibitemOpen
  \bibfield  {author} {\bibinfo {author} {\bibfnamefont {N.~N.}\ \bibnamefont
  {Bogoliubov}},\ }\href@noop {} {\emph {\bibinfo {title} {Problems of Dynamic
  Theory in Statistical Physics}}}\ (\bibinfo  {publisher} {Studies in
  Statistical Mechanics, Vol. I, North-Holland, Amsterdam},\ \bibinfo {year}
  {1962})\BibitemShut {NoStop}%
\bibitem [{\citenamefont {Boltzmann}(1964)}]{boltzmann1964lectures}%
  \BibitemOpen
  \bibfield  {author} {\bibinfo {author} {\bibfnamefont {L.}~\bibnamefont
  {Boltzmann}},\ }\href@noop {} {\emph {\bibinfo {title} {Lectures on gas
  theory}}}\ (\bibinfo  {publisher} {University of California Press},\ \bibinfo
  {year} {1964})\BibitemShut {NoStop}%
\bibitem [{\citenamefont {Ehrenfest}\ and\ \citenamefont
  {Ehrenfest}(1990)}]{ehrenfest1990conceptual}%
  \BibitemOpen
  \bibfield  {author} {\bibinfo {author} {\bibfnamefont {P.}~\bibnamefont
  {Ehrenfest}}\ and\ \bibinfo {author} {\bibfnamefont {T.}~\bibnamefont
  {Ehrenfest}},\ }\href@noop {} {\emph {\bibinfo {title} {The conceptual
  foundations of the statistical approach in mechanics}}}\ (\bibinfo
  {publisher} {Courier Corporation},\ \bibinfo {year} {1990})\BibitemShut
  {NoStop}%
\bibitem [{\citenamefont {Brown}\ \emph {et~al.}(2009)\citenamefont {Brown},
  \citenamefont {Myrvold},\ and\ \citenamefont {Uffink}}]{Brown2009}%
  \BibitemOpen
  \bibfield  {author} {\bibinfo {author} {\bibfnamefont {H.~R.}\ \bibnamefont
  {Brown}}, \bibinfo {author} {\bibfnamefont {W.}~\bibnamefont {Myrvold}}, \
  and\ \bibinfo {author} {\bibfnamefont {J.}~\bibnamefont {Uffink}},\ }\href
  {\doibase https://doi.org/10.1016/j.shpsb.2009.03.003} {\bibfield  {journal}
  {\bibinfo  {journal} {Stud. Hist. Philos. M. P.}\ }\textbf {\bibinfo {volume}
  {40}},\ \bibinfo {pages} {174} (\bibinfo {year} {2009})}\BibitemShut
  {NoStop}%
\bibitem [{\citenamefont {Wegner}(1980)}]{Wegner1980}%
  \BibitemOpen
  \bibfield  {author} {\bibinfo {author} {\bibfnamefont {F.}~\bibnamefont
  {Wegner}},\ }\href {\doibase 10.1007/BF01325284} {\bibfield  {journal}
  {\bibinfo  {journal} {Z. Phys. B Condens. Matter}\ }\textbf {\bibinfo
  {volume} {36}},\ \bibinfo {pages} {209} (\bibinfo {year} {1980})}\BibitemShut
  {NoStop}%
\bibitem [{\citenamefont {Fu}\ \emph {et~al.}(2019{\natexlab{a}})\citenamefont
  {Fu}, \citenamefont {Zhang},\ and\ \citenamefont
  {Zhao}}]{PhysRevE.100.010101}%
  \BibitemOpen
  \bibfield  {author} {\bibinfo {author} {\bibfnamefont {W.}~\bibnamefont
  {Fu}}, \bibinfo {author} {\bibfnamefont {Y.}~\bibnamefont {Zhang}}, \ and\
  \bibinfo {author} {\bibfnamefont {H.}~\bibnamefont {Zhao}},\ }\href {\doibase
  10.1103/PhysRevE.100.010101} {\bibfield  {journal} {\bibinfo  {journal}
  {Phys. Rev. E}\ }\textbf {\bibinfo {volume} {100}},\ \bibinfo {pages}
  {010101} (\bibinfo {year} {2019}{\natexlab{a}})}\BibitemShut {NoStop}%
\bibitem [{\citenamefont {Fu}\ \emph {et~al.}(2019{\natexlab{b}})\citenamefont
  {Fu}, \citenamefont {Zhang},\ and\ \citenamefont {Zhao}}]{Fu_2019}%
  \BibitemOpen
  \bibfield  {author} {\bibinfo {author} {\bibfnamefont {W.}~\bibnamefont
  {Fu}}, \bibinfo {author} {\bibfnamefont {Y.}~\bibnamefont {Zhang}}, \ and\
  \bibinfo {author} {\bibfnamefont {H.}~\bibnamefont {Zhao}},\ }\href {\doibase
  10.1088/1367-2630/ab115a} {\bibfield  {journal} {\bibinfo  {journal} {New
  Journal of Physics}\ }\textbf {\bibinfo {volume} {21}},\ \bibinfo {pages}
  {043009} (\bibinfo {year} {2019}{\natexlab{b}})}\BibitemShut {NoStop}%
\bibitem [{\citenamefont {Fu}\ \emph {et~al.}(2019{\natexlab{c}})\citenamefont
  {Fu}, \citenamefont {Zhang},\ and\ \citenamefont
  {Zhao}}]{PhysRevE.100.052102}%
  \BibitemOpen
  \bibfield  {author} {\bibinfo {author} {\bibfnamefont {W.}~\bibnamefont
  {Fu}}, \bibinfo {author} {\bibfnamefont {Y.}~\bibnamefont {Zhang}}, \ and\
  \bibinfo {author} {\bibfnamefont {H.}~\bibnamefont {Zhao}},\ }\href {\doibase
  10.1103/PhysRevE.100.052102} {\bibfield  {journal} {\bibinfo  {journal}
  {Phys. Rev. E}\ }\textbf {\bibinfo {volume} {100}},\ \bibinfo {pages}
  {052102} (\bibinfo {year} {2019}{\natexlab{c}})}\BibitemShut {NoStop}%
\bibitem [{\citenamefont {Pistone}\ \emph {et~al.}(2019)\citenamefont
  {Pistone}, \citenamefont {Chibbaro}, \citenamefont {Bustamante},
  \citenamefont {Lvov},\ and\ \citenamefont {Onorato}}]{Onorato2019}%
  \BibitemOpen
  \bibfield  {author} {\bibinfo {author} {\bibfnamefont {L.}~\bibnamefont
  {Pistone}}, \bibinfo {author} {\bibfnamefont {S.}~\bibnamefont {Chibbaro}},
  \bibinfo {author} {\bibfnamefont {M.~D.}\ \bibnamefont {Bustamante}},
  \bibinfo {author} {\bibfnamefont {Y.~V.}\ \bibnamefont {Lvov}}, \ and\
  \bibinfo {author} {\bibfnamefont {M.}~\bibnamefont {Onorato}},\ }\href
  {\doibase 10.3934/mine.2019.4.672} {\bibfield  {journal} {\bibinfo  {journal}
  {Math. Eng.}\ }\textbf {\bibinfo {volume} {1}},\ \bibinfo {pages} {672}
  (\bibinfo {year} {2019})}\BibitemShut {NoStop}%
\bibitem [{\citenamefont {Wang}\ \emph {et~al.}(2020)\citenamefont {Wang},
  \citenamefont {Fu}, \citenamefont {Zhang},\ and\ \citenamefont
  {Zhao}}]{PhysRevLett.124.186401}%
  \BibitemOpen
  \bibfield  {author} {\bibinfo {author} {\bibfnamefont {Z.}~\bibnamefont
  {Wang}}, \bibinfo {author} {\bibfnamefont {W.}~\bibnamefont {Fu}}, \bibinfo
  {author} {\bibfnamefont {Y.}~\bibnamefont {Zhang}}, \ and\ \bibinfo {author}
  {\bibfnamefont {H.}~\bibnamefont {Zhao}},\ }\href {\doibase
  10.1103/PhysRevLett.124.186401} {\bibfield  {journal} {\bibinfo  {journal}
  {Phys. Rev. Lett.}\ }\textbf {\bibinfo {volume} {124}},\ \bibinfo {pages}
  {186401} (\bibinfo {year} {2020})}\BibitemShut {NoStop}%
\bibitem [{\citenamefont {Fu}\ \emph {et~al.}(2021)\citenamefont {Fu},
  \citenamefont {Zhang},\ and\ \citenamefont {Zhao}}]{PhysRevE.104.L032104}%
  \BibitemOpen
  \bibfield  {author} {\bibinfo {author} {\bibfnamefont {W.}~\bibnamefont
  {Fu}}, \bibinfo {author} {\bibfnamefont {Y.}~\bibnamefont {Zhang}}, \ and\
  \bibinfo {author} {\bibfnamefont {H.}~\bibnamefont {Zhao}},\ }\href {\doibase
  10.1103/PhysRevE.104.L032104} {\bibfield  {journal} {\bibinfo  {journal}
  {Phys. Rev. E}\ }\textbf {\bibinfo {volume} {104}},\ \bibinfo {pages}
  {L032104} (\bibinfo {year} {2021})}\BibitemShut {NoStop}%
\bibitem [{\citenamefont {Feng}\ \emph {et~al.}(2022)\citenamefont {Feng},
  \citenamefont {Fu}, \citenamefont {Zhang},\ and\ \citenamefont
  {Zhao}}]{Feng_2022}%
  \BibitemOpen
  \bibfield  {author} {\bibinfo {author} {\bibfnamefont {S.}~\bibnamefont
  {Feng}}, \bibinfo {author} {\bibfnamefont {W.}~\bibnamefont {Fu}}, \bibinfo
  {author} {\bibfnamefont {Y.}~\bibnamefont {Zhang}}, \ and\ \bibinfo {author}
  {\bibfnamefont {H.}~\bibnamefont {Zhao}},\ }\href {\doibase
  10.1088/1742-5468/ac6a5a} {\bibfield  {journal} {\bibinfo  {journal} {J.
  Stat. Mech. Theory Exp.}\ }\textbf {\bibinfo {volume} {2022}},\ \bibinfo
  {pages} {053104} (\bibinfo {year} {2022})}\BibitemShut {NoStop}%
\bibitem [{\citenamefont {Callen}\ and\ \citenamefont
  {Welton}(1951)}]{PhysRev.83.34}%
  \BibitemOpen
  \bibfield  {author} {\bibinfo {author} {\bibfnamefont {H.~B.}\ \bibnamefont
  {Callen}}\ and\ \bibinfo {author} {\bibfnamefont {T.~A.}\ \bibnamefont
  {Welton}},\ }\href {\doibase 10.1103/PhysRev.83.34} {\bibfield  {journal}
  {\bibinfo  {journal} {Phys. Rev.}\ }\textbf {\bibinfo {volume} {83}},\
  \bibinfo {pages} {34} (\bibinfo {year} {1951})}\BibitemShut {NoStop}%
\bibitem [{\citenamefont {Margalit}\ and\ \citenamefont
  {Rabino}(2019)}]{LinearAlgebra2019}%
  \BibitemOpen
  \bibfield  {author} {\bibinfo {author} {\bibfnamefont {D.}~\bibnamefont
  {Margalit}}\ and\ \bibinfo {author} {\bibfnamefont {J.}~\bibnamefont
  {Rabino}},\ }\href@noop {} {\emph {\bibinfo {title} {Interactive Linear
  Algebra}}}\ (\bibinfo  {publisher} {Georgia Institute of Technology},\
  \bibinfo {year} {2019})\BibitemShut {NoStop}%
\bibitem [{\citenamefont {Xiong}\ \emph {et~al.}(2012)\citenamefont {Xiong},
  \citenamefont {Wang}, \citenamefont {Zhang},\ and\ \citenamefont
  {Zhao}}]{PhysRevE.85.020102}%
  \BibitemOpen
  \bibfield  {author} {\bibinfo {author} {\bibfnamefont {D.}~\bibnamefont
  {Xiong}}, \bibinfo {author} {\bibfnamefont {J.}~\bibnamefont {Wang}},
  \bibinfo {author} {\bibfnamefont {Y.}~\bibnamefont {Zhang}}, \ and\ \bibinfo
  {author} {\bibfnamefont {H.}~\bibnamefont {Zhao}},\ }\href {\doibase
  10.1103/PhysRevE.85.020102} {\bibfield  {journal} {\bibinfo  {journal} {Phys.
  Rev. E}\ }\textbf {\bibinfo {volume} {85}},\ \bibinfo {pages} {020102}
  (\bibinfo {year} {2012})}\BibitemShut {NoStop}%
\bibitem [{\citenamefont {Popkov}\ \emph {et~al.}(2015)\citenamefont {Popkov},
  \citenamefont {Schadschneider}, \citenamefont {Schmidt},\ and\ \citenamefont
  {Schütz}}]{pnas2015}%
  \BibitemOpen
  \bibfield  {author} {\bibinfo {author} {\bibfnamefont {V.}~\bibnamefont
  {Popkov}}, \bibinfo {author} {\bibfnamefont {A.}~\bibnamefont
  {Schadschneider}}, \bibinfo {author} {\bibfnamefont {J.}~\bibnamefont
  {Schmidt}}, \ and\ \bibinfo {author} {\bibfnamefont {G.~M.}\ \bibnamefont
  {Schütz}},\ }\href {\doibase 10.1073/pnas.1512261112} {\bibfield  {journal}
  {\bibinfo  {journal} {Proc. Natl. Acad. Sci.}\ }\textbf {\bibinfo {volume}
  {112}},\ \bibinfo {pages} {12645} (\bibinfo {year} {2015})}\BibitemShut
  {NoStop}%
\bibitem [{\citenamefont {Xiong}(2018)}]{PhysRevE.97.022116}%
  \BibitemOpen
  \bibfield  {author} {\bibinfo {author} {\bibfnamefont {D.}~\bibnamefont
  {Xiong}},\ }\href {\doibase 10.1103/PhysRevE.97.022116} {\bibfield  {journal}
  {\bibinfo  {journal} {Phys. Rev. E}\ }\textbf {\bibinfo {volume} {97}},\
  \bibinfo {pages} {022116} (\bibinfo {year} {2018})}\BibitemShut {NoStop}%
\bibitem [{\citenamefont {Chen}\ \emph {et~al.}(2015)\citenamefont {Chen},
  \citenamefont {Wang}, \citenamefont {Casati},\ and\ \citenamefont
  {Benenti}}]{Chen2015}%
  \BibitemOpen
  \bibfield  {author} {\bibinfo {author} {\bibfnamefont {S.}~\bibnamefont
  {Chen}}, \bibinfo {author} {\bibfnamefont {J.}~\bibnamefont {Wang}}, \bibinfo
  {author} {\bibfnamefont {G.}~\bibnamefont {Casati}}, \ and\ \bibinfo {author}
  {\bibfnamefont {G.}~\bibnamefont {Benenti}},\ }\href {\doibase
  10.1103/PhysRevE.92.032139} {\bibfield  {journal} {\bibinfo  {journal} {Phys.
  Rev. E}\ }\textbf {\bibinfo {volume} {92}},\ \bibinfo {pages} {032139}
  (\bibinfo {year} {2015})}\BibitemShut {NoStop}%
\bibitem [{\citenamefont {Zhong}\ \emph {et~al.}(2012)\citenamefont {Zhong},
  \citenamefont {Zhang}, \citenamefont {Wang},\ and\ \citenamefont
  {Zhao}}]{PhysRevE.85.060102}%
  \BibitemOpen
  \bibfield  {author} {\bibinfo {author} {\bibfnamefont {Y.}~\bibnamefont
  {Zhong}}, \bibinfo {author} {\bibfnamefont {Y.}~\bibnamefont {Zhang}},
  \bibinfo {author} {\bibfnamefont {J.}~\bibnamefont {Wang}}, \ and\ \bibinfo
  {author} {\bibfnamefont {H.}~\bibnamefont {Zhao}},\ }\href {\doibase
  10.1103/PhysRevE.85.060102} {\bibfield  {journal} {\bibinfo  {journal} {Phys.
  Rev. E}\ }\textbf {\bibinfo {volume} {85}},\ \bibinfo {pages} {060102}
  (\bibinfo {year} {2012})}\BibitemShut {NoStop}%
\bibitem [{\citenamefont {Benettin}\ \emph {et~al.}(2023)\citenamefont
  {Benettin}, \citenamefont {Orsatti},\ and\ \citenamefont
  {Ponno}}]{Benettin2023}%
  \BibitemOpen
  \bibfield  {author} {\bibinfo {author} {\bibfnamefont {G.}~\bibnamefont
  {Benettin}}, \bibinfo {author} {\bibfnamefont {G.}~\bibnamefont {Orsatti}}, \
  and\ \bibinfo {author} {\bibfnamefont {A.}~\bibnamefont {Ponno}},\ }\href
  {\doibase 10.1007/s10955-023-03147-x} {\bibfield  {journal} {\bibinfo
  {journal} {Journal of Statistical Physics}\ }\textbf {\bibinfo {volume}
  {190}},\ \bibinfo {pages} {131} (\bibinfo {year} {2023})}\BibitemShut
  {NoStop}%
\end{thebibliography}%

\end{document}